\documentclass[twocolumn,acs,longbibliography]{revtex4-1}
\usepackage{graphicx}
\usepackage{amsmath}
\usepackage{latexsym}
\usepackage{amsmath}
\usepackage{amssymb}
\usepackage{float}
\usepackage{color}
\usepackage{epstopdf}
\usepackage{hyperref}
\hypersetup{urlcolor=red,citecolor=blue}
\usepackage{bm}
\hypersetup{urlcolor=red,citecolor=blue}

\usepackage{epsfig}

\newcommand{\xmat}{\underline{\underline{X}}}

\newcommand{\gmat}{\underline{\underline{\Gamma}}}

\newcommand{\vk}{\vec{k}}
\hypersetup{urlcolor=red,citecolor=blue}

\begin{document}
\title{Partitioning of total charge in matter from geometric phases of electrons} 
\author{Joyeta Saha, Sujith Nedungattil Subrahmanian, Joydeep Bhattacharjee\\
\small{\textit{School of Physical Sciences}}\\
\small{\textit{National Institute of Science Education and Research,\\ 
A CI of Homi Bhabha National Institute, Odisha - 752050, India}}}

\begin{abstract}
Based on geometric phases of Bloch electrons computed from first-principles,
we propose a scheme for unambiguous partitioning of charge in matter, 
derivable directly from the Kohn-Sham states.
Generalizing the fact that geometric phases acquired by electrons due to evolution of their crystal momentum $\vk $ in a direction through out the Brillouin zone(BZ), provide position of their localization with net minimum spread along the 
corresponding direction in real space, 
We find that the total charge can be meaningfully distributed into charge centres
simultaneously contributed by triads of electrons with their crystal momentum evolving linearly independently 
through each unique $\vk$ across the BZ. 
The resultant map of charge centres readily renders 
not only the qualitative nature of inter-atomic as well as intra-atomic hybridization of electrons,
but also unbiased quantitative estimates of electrons on atoms or shared between them, 
as demonstrated in a select variety of isolated and 
periodic systems with varying degree of sharing of valence electrons
among atoms, including variants of multi-centered bonds.
\end{abstract}

\maketitle
\section{Introduction}
In a system of atoms, partitioning the total charge into amounts which can 
be exclusively associated with one or an irreducibly minimal number of atoms,
and deducing a picture of chemical bonding therefrom\cite{mulliken1967spectroscopy}
\cite{bader1971partitioning},
has been an ever maturing quest in pursuit of exactitude, 
since we began computing the ground states of materials more than half a century ago, 
augmented in particular by the advent of the density functional theory \cite{kohn1965self}. 
With the advent of new classes of technologically relevant materials, the debate about the nature of covalent interaction
in such systems beyond the two centre two electron picture, 
has rekindled interest in determining the true nature of sharing of electrons where multiple .
A plethora of techniques have been evolved over the years in this direction,
following primarily two thematic approaches.
In likely the more popular one, topological analysis is performed of a suitable scalar field, 
typically a component of total energy, 
and the nature of chemical bonding is
deduced from the distribution of critical points\cite{bader1985atoms,silvi1994classification}.

Techniques in the other direction 
aim to distribute the total charge into population of spatially localized orbitals 
which can be associated with individual atoms or a minimal combinations of them. 
This approach thus calls for rational construction of localized orbitals from the electronic structure of the ground state.
Construction of localized orbitals of different variants proposed so far can be broadly categorized in three directions - 
1. based on 
maximization of overlap within smaller sub-groups of orbitals
\cite{foster1980natural,krapp2006orbital,glendening2012natural},
2.  minimization of a specific localization criteria like Coulomb repulsion among orbitals
\cite{boys1960construction,edmiston1963localized,pipek1989fast} or Mulliken charge on neighboring atoms\cite{pipek1989fast}, 
and 3.  explicit minimization of total spread of orbitals
\cite{marzari1997maximally,marzari2012maximally}.
These methods generally involve iterative numerical optimization of bound orbitals in real space.  
Explicit construction of  localized orbitals 
is thus often not only a computationally cumbersome precursor to analysis of chemical bonding, 
but also a potential source of bias due to technicalities inherent to the methods of construction.
On this backdrop, the goal of this work is to propose an approach to discern the nature of chemical bonding from
the perspective of charge centres of an unbiased collection of possible occupied  orbitals without having to construct them explicitly.

For isolated systems,
the total spread of a finite set of orthonormal orbitals to be minimum in a particular direction,
the orbitals must diagonalize the component of position operator along the direction of localization\cite{resta1994macroscopic,bhattacharjee2005geometric}. 
However, position operators corresponding to three linearly independent  directions  
do not necessarily commute in a  finite basis of orthonormal bound states.
Simultaneous maximum localization in all three directions thus will be generally not possible unless 
facilitated by symmetry of the system.
Nonetheless, in isolated systems,
maximal simultaneous diagonalization\cite{cardoso1996jacobi} of non-commuting  position operators 
in the finite basis of the energy eigenstates, 
has been shown to lead to orbitals which are
maximally localized in all three directions\cite{gygi2003computation,nacbar2014simple}.
%
However, these methods are not readily usable in periodic systems,
owing to the unbound nature of their energy eigen states, namely, the Bloch functions(BF).
Instead,  we take recourse to  Wannier functions (WF) 
\cite{wannier1937structure}
which are Fourier transform of BFs in $\vk$ space,
and thus in effect are linear combinations of BFs.
and likewise, are periodic in the the 
Born-von Karman(BvK) super-cell, 
conceived in principle to be made of an infinitely large number of unit-cells.
%
With appropriate choice of gauge for Bloch functions, 
WFs can be constructed to exponentially localize
\cite{marzari1997maximally,brouder2007exponential}
primarily within an unit-cell, thus rendering in effect a bound function. 
%

However, 
a unique analytic choice of gauge leading to maximum localization of WFs 
is possible for periodic systems only in one arbitrary direction\cite{marzari1997maximally,sgiarovello2001electron,bhattacharjee2005geometric}.
The gauge is derivable from the eigen-structure of the non-Abelian matrix generalization ($\gmat$) of the
geometric phases acquired by the Bloch states, 
as they are evolved through the first Brillouin zone(BZ) in the direction reciprocal to that of localization in real space.
$\gmat$ and thus the gauge are function of the components  of $\vk$ 
which remain invariant during the evolution of $\vk$.
Resultant WFs
are localized in one direction and  
Bloch like in the other two, and referred as
the hermaphrodite WFs\cite{sgiarovello2001electron} 
and more recently as the hybrid WFs\cite{gresch2017z2pack}. 
%
Random phases arising with every set of hybrid WFs prohibits straight forward
generalization of application of the analytic choice across the BZ and for
localization simultaneously in more than one direction. 
%
%
%
In this work we circumvent explicit construction of WFs, and show that maximal joint diagonalization of the $\gmat$ 
matrices at all $\vk$ yields 
a non-unique yet unambiguous spatial distribution of charge centres, contributed jointly by 
Bloch electrons. 
We demonstrate construction and interpretation of the proposed charge centres in 
a representative variety of covalent, partially covalent and metavalent systems.

\section{Methodology}
As already discussed, for an isolated system, construction of a set of orthonormal states with minima of total spread along the $\vec{x}$-direction 
in the basis of a finite set of eigen states $\left\{ \phi_{n=1,..,N} \right\}$, 
would amount to diagonalization of $\xmat$ where $X_{i,j}=\langle \phi_i \mid \hat{x} \mid \phi_j \rangle $.
For periodic systems, in the limit of infinitely large BvK cell, 
$\xmat$ can be evaluated in the basis of Wannier functions(WF):
%
$X_{mn}+R_x \delta_{mn}=\langle W_{m,\vec{R}} \mid \hat{x} \mid W_{n,\vec{R}} \rangle$, 
where $\mid W_{j,R} \rangle$ is $j$-th WF localized in the unit-cell $\vec{R}$.  
%
Expanding WFs in terms of Bloch functions, 
the expression leads to: 
%
\begin{eqnarray}
X_{mn}&=&i \frac{a_x}{2\pi} \int_{BZ}\langle u_{\vec{k},m} \mid \frac{\partial}{\partial k_x} \mid u_{\vec{k},n} \rangle d^3 k
\label{xx}\\
&=&
\frac{a_x}{2\pi} \int \Gamma^x_{mn}(k_y,k_z) 
dk_y dk_z \nonumber,
\end{eqnarray}
where $u_{\vec{k}j}$ are cell-periodic part of Bloch functions, and $a_x$ being the lattice constant in $x$-direction.
%
%
Generalizing for direction $\alpha$ parallel to lattice vector $\vec{a}_\alpha$ in real space, 
the derivative of $\mid u_{\vec{k},n} \rangle$ in Eqn.(\ref{xx}) can be approximated
by retaining up to the first order 
in the Taylor expansion of $u_{\vk +\Delta \vk_\alpha}$, leading to
\begin{eqnarray}
& &\xmat^\alpha 
= i \frac{|\vec{a}_\alpha|}{2\pi}
\sum_{\vk_{\beta}}  \log  \Pi_{\vk_\alpha} 
\underline{\underline{O}}^\alpha(\vk_\alpha,\vk_\beta),
\label{xalphapar}\\
\mbox{with } 
& &O^\alpha_{mn}(\vk_\alpha,\vk_\beta)=
\langle u_{\vk_\beta+\vk_\alpha,m} \mid u_{\vk_\beta + \vk_\alpha + \Delta \vk_\alpha,n} \rangle 
\label{omat},
\end{eqnarray}
where $\vk=\vk_\alpha + \vk_\beta$, $\Delta \vk_\alpha= \vec{b}_\alpha/N_\alpha$,
and $\{\vk_\alpha \}\equiv \{l\Delta  \vk_\alpha$, $0\le l<N_\alpha \}$,
$N_\alpha$ being the number of unit-cells considered in the BvK super-cell along $\vec{a}_\alpha$,
while $ \vk_\beta $ spans the other two reciprocal lattice vectors $\{\vec{b}_{\beta \ne \alpha} \}$.
This allows us to define $\xmat^\alpha(\vk_\beta)=
\log  \Pi_{\vk_\alpha} 
\underline{\underline{O}}^\alpha(\vk_\alpha,\vk_\beta)$ following (\ref{xalphapar}), since
different sets of hybrid WFs with maximum localization along $\hat{\alpha}$ can be formed for each $\vk_\beta$.
The product $\Pi_{\vk_\alpha}$ in (\ref{xalphapar}) spans $N_\alpha$ allowed $\vk_\alpha$
values starting in principle from any arbitrary value of  $\vk_\alpha$, with no impact on the resultant hybrid WFs.
We note however that the phases in the off-diagonal elements of $\xmat^\alpha(\vk_\beta)$ would depend arbitrarily 
on the starting value of $\vk_\alpha$.
It is therefore imperative to consider  $\xmat^\alpha$ to be a function of $\vk$, as:
\begin{equation}
\xmat^\alpha(\vk)\equiv\xmat^\alpha(\vk_\alpha,\vk_\beta)=
\log  \Pi^{\vk_\alpha + \vec{G}_\alpha}_{\vk_{\alpha'}=\vk_\alpha} 
\underline{\underline{O}}^\alpha(\vk_{\alpha'},\vk_\beta)
\label{xmatk}
\end{equation}
%
%
Considering all three directions in real space - parallel to the three lattice vectors,
we would therefore have three matrices at every $\vk$, namely,  $\xmat^1(\vk),\xmat^2(\vk),\xmat^3(\vk)$,
which would not necessarily commute as per the general nature of position operators projected into a finite subspace.
For isolated systems, 
(\ref{xalphapar}) implies:
\begin{equation}
 X^\alpha_{mn}= i \frac{|\vec{a}_\alpha|}{2\pi}
 \log  \int u^\star_{m}(\vec{r})  e^{-i \vec{b}_\alpha.\vec{r}} u_{n}(\vec{r}) d^3r 
\label{xiso}
\end{equation}
which is the generalization of single point Berry phase for 
a manifold of states.
%
We use the Quantum Espresso (QE) code to compute the KS-single particle state $\{ u_{\vk,n}\}$ within the Perdew-Burke-Erzenhof (PBE) exchange-correlation and construct $\{\xmat^\alpha(\vk)\}$
for the occupied manifold.

Motivated by the fact that, in isolated systems, 
the collection of expectation values $\{d^{\alpha=1..3}_{i=1..N} \}$
rendered by the common eigen-structure obtained through maximal joint diagonalization of the three matrices $\{\xmat^\alpha\}$,
directly offers charge centres
of the maximally localized Wannier functions (MLWF)
without having to construct them, 
we resort to the same 
with  $\{\xmat^\alpha(\vk)\}$ matrices at all $\vk$ through out the BZ.
%
The joint eigen-structure is computed using an iterative scheme 
which is an extension of the Jacobi method of matrix diagonalization,
where off-diagonal elements of a matrix are set to zero through rotation of coordinates by an analytic choice of angle. 
In the proposed\cite{cardoso1996jacobi} extension, 
simultaneous diagonalization of more than one square matrices is performed
irrespective of their mutual commutation, by minimizing the off-diagonal elements iteratively.
The method uses  a class of angle
of rotation leading to complex rotation matrix $U$ which would 
minimize the composite objective function involving the off-diagonal elements of $\{\xmat^\alpha\}$:
\begin{equation}
\mbox{off}(UX^1U^\dagger)+\mbox{off}(UX^2U^\dagger)+\mbox{off}(UX^3U^\dagger)
\label{obj}
\end{equation}
where $\mbox{off}(A)=\sum_{1 \le i \ne j \ge N} |A_{i j}|^2$ for an $N \times N$ matrix $A$.
$U$ is composed of all the $N(N-1)/2$ rotation matrices for each $i\ne j$ pair.
The minimization is ensured by a particular numerical choice of the 2$\times$2 block for each  $i\ne j$ pair,
detailed in Ref.\cite{cardoso1996jacobi} and summarized in Eqn.(5)-(7) in 
Ref.\cite{hossain2021hybrid}.
$U$ is updated till convergence of the objective function.
As argued in Ref.\cite{hossain2021hybrid}, minimization of the objective function, while preserving the norm, in effect implies
maximization of separation of the $i$-th and the $j$-th orbitals, akin to the Foster and Boys localization scheme\cite{foster1960canonical}.
Expectation values of the $X^\alpha(\vk)$ in the basis of the columns of the converged $U$ constitutes a location in real space - $\{d^\alpha(\vk)_{i=1...N}\}$, 
which we refer here onwards as the maximally joint Wannier centre(JWC) contributed by Bloch electrons of crystal momentum $\vk$,
in analogy to the hybrid WFs localized along $\hat{x}_\alpha$ for which the eigenvalues of $\xmat^\alpha$ are the exact WCs.

In principle JWCs contributed by Bloch electrons with one of the unique $\vk$, 
should be possible to be mapped on to those contributed from another $\vk$,
and such maps should connect all $\vk$ across the BZ.
Finding such a map is of course a conundrum.
However if we had the maps then the JWCs could be collected in sets, with each set consisting of one JWC 
contributed by each of the $\vk$, and each such set would unambiguously be associated with a single occupied orbital.  
Interestingly, for the majority of systems that we have tested with, JWCs indeed spatially group into clusters that can be easily associated with one or a few atoms in proximity. 
Location of JWCs is not unique since it intimately depends on the selection $\vk$ which in turn depends on the choice of unit-cell.
Nevertheless, the distribution of JWCs on, around, and in between atoms, unambiguously render 
the count of electrons shared and retained on atoms in bonding and non-bonding orbitals 
respectively, thus revealing the nature of chemical bonding present in the system.
%
%
%
%
We can define a distribution of the JWCs as:
\begin{equation}
\rho_{JWC}(\vec{r})= \sum^{BZ}_{\vk} \sum^N_i \delta(\vec{d}_i(\vec{k}) - \vec{r})
\label{mjwc}
\end{equation}
such that $\int\rho_{JWC} d^3r$ renders the total number of electrons.
However in this work we have demonstrated JWCs by plotting each of them explicitly forming clusters.
\subsection{Computational details}
The proposed JWCs are calculated using in-house post-processing implementation interfaced locally
with the Quantum Espresso code which is used to computes the Kohn-Sham(KS) states $\{ u_{\vk,n}\}$. 
KS states are calculated within the  Perdew-Burke-Erzenhof(PBE) exchange-correlation 
implemented in norm-conserving pseudopotentials, in a plane-wave basis of cutoff 60 Ryd.
Ground state densities are computed in plane-wave basis of 360 Ryd. 
As apparent above, JWCs are calculated in three steps.
In the first step the overlap matrix $\underline{\underline{O}}^\alpha(\vk_\alpha,\vk_\beta)$ 
as implied in (\ref{omat}) is computed 
starting with each $\vk$ in all the linearly independent directions($\alpha$) - 
$\vk_\alpha$ to $\vk_\alpha+\vec{G}_\alpha$.
Next, $\xmat^\alpha(\vk)$ is computed as implied in (\ref{xmatk}).
Finally, maximal joint diagonalization is performed for $\{\xmat^\alpha(\vk) \}$ at each $\vk$,
and the corresponding sets of expectation values $\{d^\alpha_i\}_{i=1...N}$ are calculated
and plotted as $N$ number of JWCs associated with $\vk$.   


\section{Results and Discussion}
\subsection{JWC in molecules}

\begin{figure}[t]
    \centering
    \includegraphics[width=0.8\linewidth]{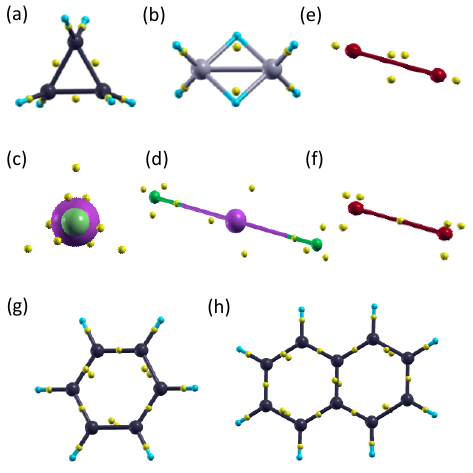}
    \caption{Distribution of spin-degenerate JWCs for (a)C$_3$H$_6$, (b)B$_2$H$_6$,
    (c)XeF$_2$(top-view), (d)XeF$_2$(side-view),
    (g)benzene, and (h)napthalene molecules.
    Spin-polarized JWCs
    of O$_2$ in 
    (e-f) for spin-1 and 2
    respectively. 
     JWCs are represented by yellow spheres.}
    \label{molecules}
\end{figure}

We exemplify distribution of JWCs first in molecules and then move to periodic systems. 
%
In isolated systems $\{\xmat^\alpha\}$ can be calculated explicitly in real space as a representation of position 
operators projected in the occupied sub-space. 
For spin polarised (spin-degenerate) calculations we associate 1 (2) electron per JWC for isolated systems.
%
In 
cyclopropane [Fig.~\ref{molecules}(a)] 
the C-C-C bond angle being $60^{\circ}$, 
does not allow orientation of hybrid atomic 
orbitals along the direction of coordination, which results into 
C-C bent-banana bonds\cite{wiberg1996bent} as evident from the deviation of the C-C JWCs 
from the direction of C-C coordination. 
In  $\textnormal{B}_2\textnormal{H}_6$ [Fig.~\ref{molecules}(b)],
JWCs near H atoms between B atoms imply three centre two electron (3c-2e)\cite{parkin2019} bond 
along each B-H-B segment. 
In principle, the JWC
near H can be considered
to contribution a total of one electron to each of the two B-H segments,
implying an effective bond order of 1/2,
known to denote electron deficiency.
%
In XeF$_2$[Fig.~\ref{molecules}(c,d)],  
JWCs suggest retention of three lone pairs on Xe and a near-$sp^3$ hybridization at F atoms, indicating
a partial ionic nature at the outset.
%
However, the four electrons represented by the JWCs on the two Xe-F segments
should in principle be shared by the F and the Xe atoms
in attempt to optimally complete their respective sub-shell filling.
A plot of the corresponding MLWFs, which is possible only for isolated
systems, suggest them to be combination of 
$5p_z(\textnormal{Xe})$ and  $2sp^3(\textnormal{F})$ orbitals,
the later being the primary component. 
These four electrons thus constitute a 4-electron 3-centre bond\cite{rundle1963}\cite{pimentel1951bonding}
where the 4 electrons are unequally shared between the Xe and F, likely
in order to avoid hypervalency of Xe,
more in sync with the charge-shift bonding\cite{braida2013essential}.
%
%
Notably, the C3 symmetry of the staggered nature of the orientation of the $sp^2$ and $sp^3$ orbitals of Xe and F, as evident from the top view[Fig.~\ref{molecules}(c)],
 is not observed if we use $\xmat^\alpha$
computed explicitly as representation of the position operator
in occupied states.
Thus the JWCs obtained using the matrix
generalization of single point Berry phase, 
can be expected to be reasonably accurate for isolated systems.
%
JWCs in {$\textnormal{O}_2$}[Fig.~\ref{molecules}(e),(f)] 
represent its di-radical nature,
with three degenerate bonding orbitals
for one spin implying a bond order of 1.5, 
and a single bonding orbital for the other spin implying bond order 0.5.
%
JWCs in benzene[Fig.~\ref{molecules}(g)]shows alternating double and single bonds although the nearest neighbor C-C distances are all same. 
This arbitrary choice of one of the resonant structures is a result of the inherent property of maximal joint diagonalization 
of $\{\xmat^\alpha\}$ to ensure maximum localization of the common eigen-states.
In naphthalene[Fig.~\ref{molecules}(h)], JWCs renders the symmetric non-degenerate distribution of the double bonds expected in the ground-state. 
%
%
%
\subsection{JWCs in periodic systems}
\begin{figure}[t]
    \centering
    \includegraphics[trim={0 3cm 0 0}]{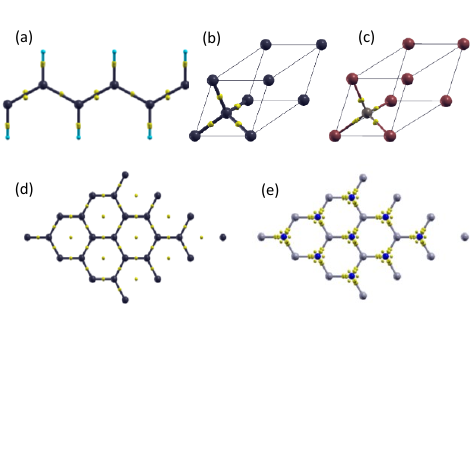}
    \caption{Unsymmetrised JWC distributions for - (a)poly-acetylene chain, (b)diamond, (c)GaAs(zinc blende). Symmetrized JWC distribution for - (d)graphene, (e)hexagonal-BN.}
    \label{Gr}
\end{figure}

In periodic systems, each JWC point represents a population of $1(2)/N_k$  electrons for
spin-polarized (spin-degenerate) ground-state, $N_k=\Pi_\alpha N_\alpha$ being the total number of unique $\vk$ in the BZ. 
%
Similar to benzene, in poly-acetylene chain[Fig.~\ref{Gr}(a)] as well, 
JWCs randomly chooses one of the resonant configurations to represent. 
Similar to the poly-acetylene chain which is a known example in one dimension where JWCs are exact WCs,
in higher dimension as well,
 JWC will choose one of the resonant configurations 
along each of the directions of evaluation of Berry phase, such that the JWCs 
are maximally apart. 
Symmetry of the distribution of JWCs will therefore follow the symmetry of the
k-paths along which $\{\xmat^\alpha\}$ are calculated.
There are therefore two reasons 
for JWCs to be non-unique - one is the random choice of a resonant configuration and the second is the dependence on the choice of k-paths which is connected to the choice of unit-cell.
%
JWCs should therefore ideally need to be symmetrized by applying the symmetries of the crystal structure exhaustively, such that the net distribution of JWCs consists of contributions from 
at least one complete set of symmetry equivalent k-paths. 
%
%
However in general, the robust features of orbital occupation 
represented by the location of JWC clusters around atoms, 
for example, the tetrahedral arrangement of four clusters of JWCs around atoms 
to denote an $sp^3$ nature of hybridization of its orbitals, 
or the two center bonds $\sigma$-bonds, 
will be discernible without symmetrization and irrespective of the choice of unit-cell with
reasonable certainty.
In covalent semiconductors and insulators 
[Fig.\ref{Gr}(b-c)]
clusters of JWCs thus duly represent homo-polar or hetero-polar $\sigma$ bonds.
As discussed, the exact shape of each JWC cluster of course differ with the choice of unit cells,
 for example the C3 symmetry of shape around 111 in bulk GaAs [Fig.\ref{Gr}(c)]
 due to rhombohedral choice of unit-cell, 
but the average location of each cluster remains same.
In case of homopolar bonds, the size of JWC clusters are much smaller, almost converging to a point in the direction of coordination, like in diamond or graphene[Fig.~\ref{Gr}(b),(d)], compared to a spreaded out cluster in the case of the Ga-As or B-N[Fig.~\ref{Gr}(c),(e)] coordination. 
%

In graphene[Fig.\ref{Gr}(d)], after symmetrization, JWCs 
on C-C render about 1.313 electrons each. 
%
The rest of about 0.06e is centered in the middle of each ring.
This population is associated with the sparse occupation of WFs due to hybridization of the diagonally opposite third neighboring $p_z$ orbitals.
We have verified the existence of similar populations as the Mayer bond order between third neighboring C atoms in graphene
in the basis of $p_z$ orbitals constructed from the same KS states used in calculation of JWCs. 
In hexagonal-BN the three JWCs off-centered away from N are indicative of the back donation of $sp$ hybridized lone-pairs of N towards B.
Notably, the difference between JWC distribution before and after symmetrization is nominal in graphene, compared to that  
in case of hexagonal-BN.
Graphene being semimetals, we chose to deliberately avoid the Dirac points, and converged the JWC distribution in terms of such k-mesh.
For metals, disentaglement of bands will be
the necessary precursor to calculation of JWCs,
envisaged as a future work.

%
%
%
%
%

%
%
%

\begin{figure}[t]
    \centering
    \includegraphics[trim={0 2cm 0 0}]{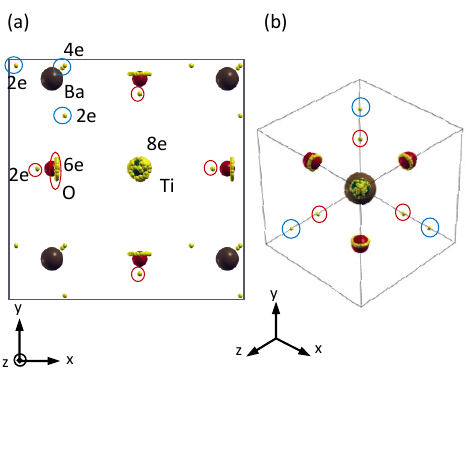}
    \caption{Unsymmetrized JWC distribution in cubic $\textnormal{BaTiO}_3$ computed with semi-core states in a 
    6$\times$6$\times$6 mesh of $\vk$.}
    \label{BTO}
\end{figure}

Next we focus on systems with partial or
more exotic forms of sharing of electrons among atoms.
In $\textnormal{BaTiO}_3$ [Fig.~\ref{BTO}]
covalent interactions are present on the dominant backdrop of ionic bonding, 
the primary covalent mechanism being the hybridization of 2$p$(O) orbitals with the 3$d$(Ti) orbitals on its two sides.
Count of JWCs around Ba, Ti and O suggests +2, +4 and -2 states as expected. 
Like in polyacetylene, without explicit symmetrization, JWCs inherently choose to describe 
one of the two resonance configurations, as evident from the asymmetry of the distribution of JWCs about O along the -Ti-O-Ti- bridge.
The ring-like distribution of  JWCs on one side of O in the -Ti-O-Ti- bridge along $\hat{\alpha}$, 
is due to hybridization of one of the two 2$sp_\alpha$(O)+3$d_{eg\parallel\alpha}$(Ti) and all possible 
linear combinations of 2$p_{\beta\ne\alpha}$(O)+3$d_{t2g\perp\alpha}$(Ti) orbitals.
The lone JWC on the other side of O is due to the other 2$sp_\alpha$(O)+3$d_{eg\parallel\alpha}$(Ti) orbital.
%
%
Tetrahedral distribution of JWCs around Ba suggests a complete sub-shell. 
The clustering of JWCs on Ti corresponds to semi-core states. 
With about 1\% displacement of Ti from the middle of the -O-Ti-O- bridge, 
Born effective charge(Z$^\star$) of Ti estimated from JWCs computed for a 
6$\times$6$\times$6 $\vk$ mesh, is found to be about 6.7a.u.
\cite{bhattacharjee2010wannier}, 
amounting to anomalous charge of about 2.7a.u.
Z$^\star$ is estimated from difference of dipole moment calculated as $\int \vec{r} \rho_{JWC} d^3 r$
post and prior to displacement of Ti atom.

\begin{figure*}[t]
    \centering
    \includegraphics[width=1.0\linewidth]{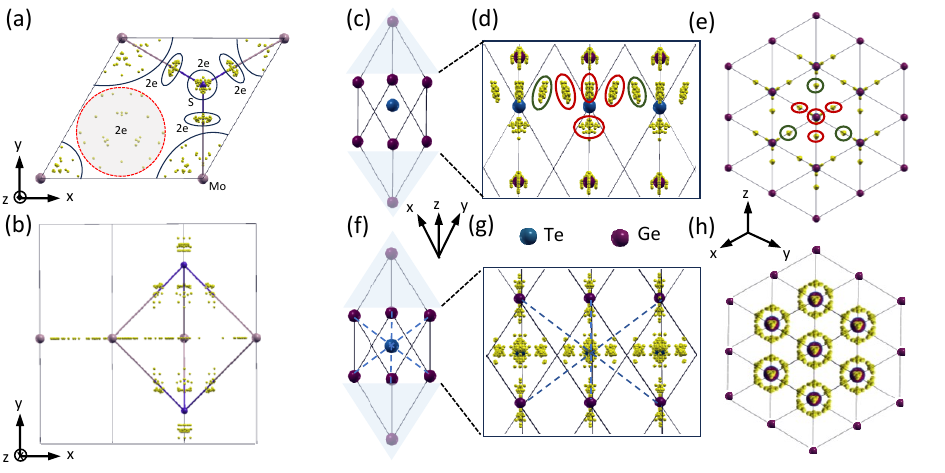}
    \caption{(a)Top and (b) side views of symmetrized JWC distribution in MoS$_2$ monolayer done with in a 6$\times$6$\times$1 mesh of $\vk$.\\
    (c) Primitive cells, (d)side and  (e) top(111) views 
    of JWC distribution in GeTe of space groups R3m 
    calculated with 6$\times$6$\times$6 k mesh.
    Same in (f-h) for symmetrized JWC distribution in GeTe of space group Fm3m(rock salt). 
    JWCS within the regions marked by the red in (d), and correspondingly in (e), are along nearest Te-Ge coordination in R3m.    
    The linear -Ge-Te-Ge segment has been marked by dashed lines in (f) and (g).}
    \label{GeTe}
\end{figure*}

Next we calculate JWCs in $\textnormal{MoS}_2$ monolayer and GeTe, 
where electrons are known to be shared by more than two atoms over 
an extended region.
The tetrahedral clustering of JWCs around S atoms in $\textnormal{MoS}_2$ monolayer [Fig.~\ref{GeTe}(b)], suggests $sp^3$ hybridization at S, 
implying sub-shell filling.
%
%
With symmetrization,
each JWC point in Fig.~\ref{GeTe}(a) 
accounts for $1/(3N_k)$ electron.
%
%
%
Notably, 
JWCs associated with Mo atoms are distributed in the Mo plane in two arrangements -
(1) group of JWCS around Mo enclosing 8 electrons, and
(2) another group (marked by the red dotted line in Fig.~\ref{GeTe}(a)) enclosing two electrons centre at the Mo interstitial not covered by S atoms, 
implying a 2-electron 3-centre covalent interaction, which is also a signature of $\sigma$ aromaticity ~\cite{Mos2} driven by Mo.


\subsubsection{Metavalent bonding in $\textnormal{GeTe}$}
Two phases of  GeTe - R3m, and rock-salt(Fm3m), known to have different levels of 
covalent interactions between Ge and Te planes.
Although the sequence of stacking of traingular lattices of Ge and Te atoms in alternate planes perpendicular to the C3 axis are same in the two phases [Fig.~\ref{GeTe}(c),(f)] ,
they differ in inter-planar spacing and intra-planar
seperation of atoms.
The Te and Ge planes are equispaced in the rock-salt(RS) structure,
leading to six nearest Ge and Te neighbors of Te and Ge atoms respectively,
while it is three for both atoms in R3m phases.
%
The exact nature of chemical bonding in these materials has been in active focus in recent years,
particularly regarding the degree of multicentredness in the Ge-Te-Ge linear segment in the RS phase,
argued from different perspectives to be 
electron rich hypervalent
\cite{hempelmann2022orbital,lee2020chemical}
, as well as electron deficient metavelent
\cite{GeTe}.
We calculate JWCs in the rhombohedral unit-cell in both cases and symmetrize the JWCs using inversion about the midpoint of the unit-cell 
in case of RS.
%
%
JWCs suggest Ge to be electron deficient with two electrons, while there are eight electrons centered around Te. 
The distribution of JWCs around Te is 
substantially different in two cases.
%
In the R3m phase, the tetrahedral clustering of JWCs suggests near-$sp^3$ hybridization at Te[Fig.~\ref{GeTe}(c)].
The clustering evolves from near-$sp^3$ to similar to $sp^2+p_z$ as the structure evolves 
from R3m to RS.
In RS phase the hexagonal clustering
[Fig.~\ref{GeTe}(h)] of JWCs with small buckling, is due to three near-$sp^2$ clusters and their inverted counterparts. 
Thus in RS the clusters of JWCs associated with the Te atoms are spread out in the Te plane[Fig.~\ref{GeTe}(g)], whereas in
R3m they are located between the two nearer Ge and Te planes.
In RS phase, three linear -Ge-Te-Ge- segments intersects at each Te, 
and vise-versa. 
The $sp^2$ or $sp^3$ nature of JWC clusters in the two phases 
suggests role of 
$s$ and $p$ electrons of Te as 
frontier orbitals on equal footing.
Rather, the $p_z$ electrons appear to 
remain largely centered on the Te atoms.
The deviation of JWC clusters from Ge atoms also suggests a partial
$sp$ hybridization.
The small yet systematic deviation  of $sp^2$ JWC clusters  from the Te plane 
suggests delocalization of the $sp^2$ orbitals of Te towards Ge in attempt to compensate for the lack of sub-shell filling of Ge, suggesting
weak covalent interaction along the Ge-Te-Ge segment mediated by the 
six $sp^2$ electrons of Te along with minor contributions from Ge.
To characterize such interaction as a traditional n-electron m-centre bond,
we should primarily consider  contribution of 2 electrons 
from a Te atom to each of the three Ge-Te-Ge segment, and a small secondary contribution
from Ge ($\le$ 2/6) from each of the Ge atoms, totalling to $2<n\le 2.67$,
which is closer to the 2-electron 3-centre metavalent bond proposed for the 
segment\cite{GeTe,wuttig2024metavalent}.
A parallel can in fact be drawn to 2-electron 3-centre bond seen in B$_2$H$_6$
if we consider an effective contribution of
one electron to each Ge-Te segment from the share
of two electrons associated with each Ge-Te-Ge 
segment, leading in effect to Ge-Te bond order of
1/2, denoting the metavalent nature.
%
The same however can not be claimed for R3m phase where the -Ge-Te-Ge- segment is neither linear, nor equispaced, and JWC clusters are localized between the shorter -Ge-Te- segment, implying an effective Ge-Te bond order should be close to 1. 
%
%


\section{Conclusion}
To conclude, for gapped systems,
 in this work we propose a map of charge centers of all possible 
maximally apart Wannnier functions contributed by all the unique set of Bloch electrons across the BvK supercell, derived through maximal joint diagonalization of Berry phase matrices computed at all allowed $\vk$
in the first Brillouin zone.
The charge centers, thus referred as the joint Wannier centers (JWC),
 are typically distributed around atoms or in between them in clusters.
While the shape and orientation of the JWC clusters depend on the choice of unit-cell,
 and calls for obvious symmetrization as per the crystal structure,
 their average relative orientation and charge content are rather unambiguous,
 and facilitate robust deduction of the nature of chemical bonding through 
quantification of sharing and transfer of electrons among atoms.
Symmetrized JWCs reveal all the possible channels of inter-atomic 
sharing or donation of electrons.
JWCs suggest prevalence of electron deficient metavalent bonding in cubic GeTe as recently claimed.

\section{Acknowledgement}
All calculations have been performed in computational resources funded by the 
Dept. of Atomic Energy of the Govt. of India.

\bibliography{ref}

\end{document}